\documentclass[twocolumn,showpacs,prl,
superscriptaddress,proprietresses]{revtex4}
\usepackage[dvips]{graphicx}
\usepackage{amsmath,amssymb}
\usepackage{dcolumn}
\begin{document}

\title{Causal dissipative hydrodynamics obtained from
 the nonextensive/dissipative correspondence}
\author{Takeshi Osada} \email{osada@ph.ns.musashi-tech.ac.jp}
\affiliation{Theoretical Physics Laboratory,
Faculty of Knowledge Engineering,
Musashi Institute of Technology, Setagaya-ku, Tokyo 158-8557, Japan}
\author{Grzegorz Wilk} \email{wilk@fuw.edu.pl}
\affiliation{The Andrzej So{\l}tan Institute for Nuclear Studies,
Ho\.{z}a 69, 00681, Warsaw, Poland}
\date{\today}

\begin{abstract}
We derive the constitutive equations of causal relativistic
dissipative hydrodynamics ($d$-hydrodynamics) from 
perfect nonextensive hydrodynamics ($q$-hydrodynamics) using the
nonextensive/dissipative correspondence (NexDC) proposed by us
recently. The $q$-hydrodynamics can be thus regarded as a possible
model for the $d$-hydrodynamics facilitating its application to
high energy multiparticle production processes. As an example we
have shown that applying the NexDC to the perfect $1+1$
$q$-hydrodynamics, one obtains a proper time evolution of the bulk
pressure and the Reynolds number.
\end{abstract}
\maketitle

The ideal hydrodynamic model successfully reproduces most of the
RHIC experimental data \cite{Hirano0808.2684} suggesting that
matter created there resembles a strongly interacting quark-gluon
fluid \cite{TDLeeNPA750} rather than a free parton gas. However,
there are also hints that, after all, a hadronic fluid cannot be
totally ideal \cite{nonviscousindic} and should be described by
some kind of dissipative hydrodynamic model (or $d$-hydrodynamics)
\cite{Eckart1940,HiscockPRD31,IsraelAnnPhys118}. A number of such
models have been proposed recently \cite{MurongaPRC69,
HeinzPRC73,BaierPRC73,KoidePRC75,ChaudhuriPRC74}. Although very
promising,  they have many difficulties, both in what concerns
their proper formulation and applications
\cite{HiscockPRD31,KoidePRC75,problems}. Any new attempt to
understand them more deeply is therefore welcome. In this letter
we demonstrate that one can obtain constitutive equations of
causal $d$-hydrodynamics starting from the relativistic perfect
nonextensive hydrodynamics ($q$-hydrodynamics) supplemented by the
nonextensive/dissipative correspondence (NexDC) proposed in
\cite{OsadaPRC77} and taking place between the perfect
$q$-hydrodynamics (based on the nonextensive Tsallis statistics
\cite{Tsallis} and described by the nonextensivity parameter $q$)
and the usual $d$-hydrodynamics (based on the extensive
Boltzmann-Gibbs statistics (BG) to which Tsallis statistics
converges for $q \rightarrow 1$). We argue that the
$d$-hydrodynamics emerging from such $q$-hydrodynamics is causal
and preserves the simplicity of the latter which facilitates
numerical applications \cite{OsadaPRC77} (instead of solving very
complicated second order equations of $d$-hydrodynamics one can
solve the much simpler equation of motion of the
$q$-hydrodynamics).

Our $q$-hydrodynamics \cite{OsadaPRC77} is based on the
relativistic nonextensive kinetic theory proposed in
\cite{LavagnoPhysLettA301} (see also \cite{LimaPhysRevLett86}). It
is understood that such a theory accounts, by means of the
parameter $q$, for all kinds of possible strong intrinsic
fluctuations and long-range correlations existing in a hadronizing
system. It replaces the usual notion of local thermal equilibrium
by a kind of stationary state which also includes some
interactions. It is described by the Tsallis nonextensive
statistics \cite{Tsallis} and characterized by the nonextensive
parameter $q> 1$ \cite{KodamaJPhys31,SR,Biro}. In this approach
the {\it perfect $q$-hydrodynamic equations} (i.e., equations for
the perfect nonextensive $q$-fluid without any additional
currents) are given by the $q$-version of the energy momentum
tensor:
\begin{eqnarray}
{\cal T}_{q;\mu}^{\mu\nu} = \Big[ \varepsilon_q(T_q)
u_q^{\mu}u_q^{\nu} - P_q(T_q) \Delta_q^{\mu\nu} \Big]_{;\mu} = 0.
\label{eq:q-equation_of_motion}
\end{eqnarray}
Here $\varepsilon_q(T_q)$, $P_q(T_q)$, $T_q=T_q(x)$ and
$u^{\mu}_q(x)$ are, respectively, the nonextensive energy density,
pressure and temperature field \cite{OsadaPRC77} and an
accompanying hydrodynamic $q$-flow four vector field (which allows
decomposition (\ref{eq:q-equation_of_motion}) to be performed;
$\Delta_q^{\mu\nu} \equiv g^{\mu\nu}-u_q^{\mu}u_q^{\nu}$). If
$T(x)$ and $u^{\mu}(x)$ are the corresponding temperature and
velocity fields in the case of BG statistics ($q=1$) with
$\varepsilon$ and $P$ being the usual energy density and pressure
then, as proposed in \cite{OsadaPRC77}, one can map a $q$-flow
into some dissipative $d$-flow by requiring  that the following
relations hold:
\begin{eqnarray}
P(T) = P_q(T_q),\quad \varepsilon(T) = \varepsilon_q(T_q) + 3\Pi,
\label{eq:Nex/diss}
\end{eqnarray}
where $\Pi$ is to be regarded as the bulk pressure of
$d$-hydrodynamics which is defined by
\begin{eqnarray}
\Pi  \equiv \frac{1}{3} w_q [\gamma^2+2\gamma] . \label{eq:bulk_pressure}
\end{eqnarray}
Here $\delta u_q^{\mu} \equiv u_q^{\mu}-u^{\mu}$ is the four
velocity difference between the non-extensive parameter $q>1$ and
$q=1$ flows, $\gamma(x) \equiv \delta u_q^{\mu}(x) u_{\mu}(x) =
-\frac{1}{2}\delta u_{q\mu} \delta u_q^{\mu}$ and $w_q\equiv
\varepsilon_q+P_q$ is the $q$-enthalpy. If $T(x)$ and $\gamma(x)$
satisfy Eq.~(\ref{eq:Nex/diss}) relations in the whole space-time
region, one can always transform equation of perfect
$q$-hydrodynamics Eq.~(\ref{eq:q-equation_of_motion}) into the
following equation of
$d$-hydrodynamics \cite{Eckart1940,HiscockPRD31,IsraelAnnPhys118,%
MurongaPRC69,HeinzPRC73,BaierPRC73, KoidePRC75,
ChaudhuriPRC74}:
\begin{eqnarray}
\left[ \varepsilon(T) u^{\mu}u^{\nu} \!-\!\left[ P(T)+\Pi \right]
\Delta^{\mu\nu} \!\!\!+\! 2 W^{(\mu} u^{\nu )} \!+\!\pi^{\mu\nu}
\right]_{;\mu} \!\!\!\!=0, \label{eq:Nex/diss_equation}
\end{eqnarray}
in which one recognizes the energy flow vector $W^{\mu}$ and the
(symmetric and traceless) shear pressure tensor $\pi^{\mu \nu}$,
\begin{equation}
W^{\mu} = w_q[1+\gamma] \Delta^{\mu}_{\sigma} \delta
u_q^{\sigma},\quad \pi^{\mu\nu} = w_q \delta u_q^{<\mu} \delta
u_q^{\nu >} \label{eq:Wpi}
\end{equation}
(where $\Delta^{\mu\nu} \equiv g^{\mu\nu}-u^{\mu}u^{\nu}$, $A^{(
\mu}B^{\nu )}\equiv \frac{1}{2}(A^{\mu}B^{\nu} +A^{\nu}B^{\mu})$,
and $a^{<\mu}b^{ \nu >} \equiv [\frac{1}{2}(\Delta^{\mu}_{\lambda}
\Delta^{\nu}_{\sigma} +
  \Delta^{\mu}_{\sigma} \Delta^{\nu}_{\lambda} )
  -\frac{1}{3}\Delta^{\mu\nu} \Delta_{\lambda\sigma} ]
  a^{\lambda}b^{\sigma}$). The $d$-hydrodynamics represented
by Eq.~(\ref{eq:Nex/diss_equation}) can be therefore regarded as a
viscous counterpart of the perfect $q$-hydrodynamics represented
by Eq.~(\ref{eq:q-equation_of_motion}).  We call this relation the
{\it nonextensive/dissipative correspondence}, (NexDC). With the
bulk pressure Eq.~(\ref{eq:bulk_pressure}) and NexDC relations
Eq.~(\ref{eq:Nex/diss}) one gets the $q$-enthalpy $w_q =
w/[1+\gamma]^2$ with $w\equiv Ts = \varepsilon+P$ being the usual
enthalpy in BG statistics. Accordingly, the bulk pressure
Eq.~(\ref{eq:bulk_pressure}) can be written as
\begin{eqnarray}
\Pi =  w  \Gamma,\quad {\rm where}\quad \Gamma\equiv
\frac{1}{3}\frac{\gamma(\gamma+2)}{(\gamma + 1)^2}.
\label{eq:Gamma}
\end{eqnarray}
Finally, the NexDC leads to the following relations between
components of the dissipative tensor:
\begin{equation}
W^{\mu} W_{\mu} \!=\! -3\Pi w,~ \pi^{\mu\nu}W_{\nu} \!=\! -2 \Pi
W^{\mu},~  \pi_{\mu\nu}\pi^{\mu\nu} \!=\! 6\Pi^2. \label{eq:pipi}
\end{equation}

We shall now address the central question of this Letter: is
$d$-hydrodynamics (\ref{eq:Nex/diss_equation}) obtained in this
way causal, as is the one discussed in \cite{MurongaPRC69,
HeinzPRC73,BaierPRC73,KoidePRC75,ChaudhuriPRC74}? To this end, let
us first notice that, because we expect  that $q-1\ll 1$
\cite{OsadaPRC77}, our stationary state described by
Eqs.~(\ref{eq:q-equation_of_motion}) and
(\ref{eq:Nex/diss_equation}) can be regarded as a {\it near
equilibrium state} (with energy momentum tensor ${\cal
T}^{\mu\nu}_q \equiv (\varepsilon_q+P_q) u^{\mu}_q u^{\nu}_q-P_q
g^{\mu\nu} $). Accordingly, the state with $q=1$, i.e., the one
with no residual correlations between fluid elements and no
intrinsic fluctuations present, is a {\it true equilibrium state}
(with energy momentum tensor ${\cal T}^{\mu\nu}_{\rm eq} \equiv
\varepsilon(T) u^{\mu}u^{\nu} -P(T)\Delta^{\mu\nu}$ and with the
equilibrium distribution given by the usual Boltzmann distribution
function, $f_{\rm eq}(x,p)=\exp \left[-p^{\mu}u_{\mu}(x)/k_{\rm B}
T(x) \right]$). Both states are near to each other and ${\cal
T}^{\mu\nu}_q = {\cal T}^{\mu\nu}_{\rm eq} + \delta {\cal
T}^{\mu\nu},$ where $\delta {\cal T}^{\mu\nu} =
-\Pi\Delta^{\mu\nu} + 2 W^{(\mu} u^{\nu )} + \pi^{\mu\nu}$. As an
immediate consequence of this, the conservation of $q$-entropy
assumed in the ideal $q$-hydrodynamics,  $[ s_q
u_q^{\mu}]_{;\mu}=0$, results in the production of the usual
entropy in $d$-hydrodynamics:
\begin{eqnarray}
T [s u^{\mu} ]_{;\mu} 
 \!\!&=&\!\! -\Pi u^{\mu}_{;\mu} -W^{\mu}_{;\mu} +W^{\mu}\frac{du_{\mu}}{d\tau}
 + \pi^{\mu\nu}  u_{\langle \mu;\nu \rangle}
  \label{eq:sprod}
\end{eqnarray}
($\frac{dX}{d\tau} \equiv u^{\mu} X_{;\mu}$ is the proper time
derivative acting on a quantity $X$). Acting by $d/d\tau$ on both
sides of Eq.~(\ref{eq:Gamma}) and multiplying by $\Pi$, one
obtains
\begin{subequations}
\begin{eqnarray}
&& -\Pi  \Big[ \frac{d\Pi}{d\tau}
     -  w \frac{d\Gamma}{d\tau} -  \Gamma\frac{dw}{d\tau} \Big]  \equiv 0.
     \label{eq:dPidtau}
\end{eqnarray}
Similarly, acting $d/d\tau$ on both sides of the first and the
last terms in Eq.~(\ref{eq:pipi}) one gets two other
identities:\label{eq:0identities}
\begin{eqnarray}
&&   \frac{2}{3}  \Gamma W_{\mu} \frac{dW^{\mu}}{d\tau}
     +  \Pi  \frac{d\Pi}{d\tau}
     +  \Pi  \Gamma \frac{dw}{d\tau} 
    \equiv 0,\quad
    \label{eq:dWdtau} \\
&&   -\frac{1}{6}\pi^{\mu\nu} \frac{d \pi_{\mu\nu}}{d\tau} +\Pi \frac{d\Pi}{d\tau} 
     \equiv 0.
     \label{eq:dpidtau}
\end{eqnarray}
\end{subequations}
Finally, acting by $\nabla_{\mu}$ (where $\nabla_{\mu} X^{\mu}
\equiv \Delta^{\sigma}_{\mu} X^{\mu}_{;\sigma}$) on both sides of
the middle term in  Eq.~(\ref{eq:pipi}), one gets:
\begin{eqnarray}
 \!W_{\nu} \nabla_{\mu}\pi^{\mu\nu}\! +\! \pi^{\mu\nu} \nabla_{\langle
\mu} W_{\nu\rangle}\! +\! 2W^{\mu}\nabla_{\mu}\Pi +2\Pi
\nabla_{\mu}W^{\mu}\! \equiv\! 0.\quad  \label{eq:identity_alpha}
\end{eqnarray}
Following \cite{MurongaPRC69} let us now introduce the usual
thermodynamic (positive) coefficients, $\beta_{0,1,2}$, for the,
respectively, scalar, vector and tensor dissipative contributions
to the entropy current (\ref{eq:sprod}), and the viscous/heat
coupling coefficient $\alpha$ (all with dimension [GeV]$^{-4}$ to
ensure that dimensions of identities (\ref{eq:0identities}) and
(\ref{eq:identity_alpha}) are the same as the dimension of
$T[su^{\mu}]_{;\mu}$ in Eq.~(\ref{eq:sprod})). Multiplying
(\ref{eq:dWdtau}), (\ref{eq:dpidtau}), (\ref{eq:dPidtau}) and
(\ref{eq:identity_alpha}) by the, respectively, $\beta_0$,
$\beta_1$,$\beta_2$ and $\Gamma \alpha$, and combining them
together one gets the following identity:
\begin{eqnarray}
&& \frac{2}{3} \beta_1\Gamma W_{\mu}\frac{dW^{\mu}}{d\tau} +
\alpha \Gamma W^{\nu} \nabla_{\mu}\pi^{\mu}_{\nu}  + 2
\alpha\Gamma W^{\mu} \nabla_{\mu}\Pi
\nonumber \\
&& \quad -(\beta_0-\beta_1-\beta_2)\Pi \frac{d\Pi}{d\tau} +2
\alpha \Gamma \Pi \nabla_{\mu}W^{\mu}
\nonumber \\
&& \quad -\frac{1}{6} \beta_2 \pi_{\mu\nu}
\frac{d\pi^{\mu\nu}}{d\tau} + \alpha\Gamma
\pi^{\mu\nu}\nabla_{\langle\mu}W_{\nu\rangle}
\nonumber \\
&& \quad +(\beta_0+\beta_1) \Pi\Gamma \frac{dw}{d\tau}
              + \beta_0 w \Pi \frac{d\Gamma}{d\tau}=0~.
\label{eq:NexDC_identity}
\end{eqnarray}
The $q$-entropy current is connected with $su^{\mu}$ and
$W^{\mu}/T$,
\begin{eqnarray}
 s_q u_q^{\mu} = \frac{1}{1+\gamma}
 \frac{T}{T_q}  \left( su^{\mu} +\frac{W^{\mu}}{T} \right).
\label{eq:q-entropy_current}
\end{eqnarray}
Applying the four divergence to both sides of
Eq.~(\ref{eq:q-entropy_current}) one gets (with $\phi_{\mu}\equiv
\partial_{\mu}\gamma/(1+\gamma)+\partial_{\mu}T_q/T_q$ and
$\tilde{\phi}\equiv u^{\mu}_{;\mu}-u^{\mu}\phi_{\mu}$):
\begin{eqnarray}
   W^{\mu}_{;\mu}  &=&
   - \frac{dw}{d\tau} -
     w  \tilde{\phi}  +  W^{\mu} \phi_{\mu},
\label{eq:div_heat_flow}
\end{eqnarray}
from which and Eq.~(\ref{eq:Gamma}) one has
\begin{eqnarray}
\Gamma \frac{dw}{d\tau} \!\!&=&\!\! \Gamma
(\frac{du_{\mu}}{d\tau}W^{\mu} - \nabla_{\mu}W^{\mu} ) -\Pi
\tilde{\phi} + \Gamma W^{\mu}\phi_{\mu}  \label{eq:dwdtau}
\end{eqnarray}
(with $W^{\mu}_{;\mu}=\nabla_{\mu}W^{\mu} +
u_{\mu}\frac{dW^{\mu}}{d\tau} $ and $u_{\mu} W^{\mu} =0$). Adding
now Eq.~(\ref{eq:NexDC_identity}) to (\ref{eq:sprod}) and using
Eq.~(\ref{eq:dwdtau}) one obtains
\begin{eqnarray}
T [s u^{\mu} ]_{;\mu} &=& -\Pi \Upsilon -W^{\mu}\Phi_{\mu} +
\pi^{\mu\nu}\Psi_{\mu\nu}
\end{eqnarray}
with $\Upsilon $, $\Phi_{\mu}$ and $\Psi_{\mu\nu} $ defined as
($\beta^*_0\equiv \beta_0-\beta_1$,
$\beta^*_1\equiv  \beta_1+\beta_0$,
$\alpha^*\equiv \frac{1}{\Pi\Gamma}+\beta^*_1$):
\begin{subequations}
\label{eq:GammaPhiPsi}
\begin{eqnarray}
\Upsilon \!\!&=&\!\! (\beta^*_0-\beta_2)
\frac{d\Pi}{d\tau} +u^{\mu}_{;\mu}  \nonumber \\
&+&\kappa  \beta^*_1 \tilde{\phi} \Pi -(2\alpha -\alpha^*) \Gamma
\nabla_{\mu}W^{\mu}, \label{eq:GammaPhiPsi_0}\\
\Phi_{\mu} \!\!&=&\!\!  - \frac{(\beta^*_1-\beta^*_0)}{3} \Gamma
\frac{dW_{\mu}}{d\tau}
  -(1+\alpha^*\Pi\Gamma )\frac{du_{\mu}}{d\tau}
  - \beta^*_1\Pi \Gamma \phi_{\mu} \nonumber \\
&+&\frac{(\beta^*_1+\beta^*_0)}{6} \frac{d\Gamma}{d\tau} W_{\mu}
 -\alpha \Gamma  \nabla_{\lambda} \pi^{\lambda}_{\mu} -2\alpha \Gamma \nabla_{\mu}  \Pi,\quad
 \\
\Psi_{\mu\nu}  \!\!&=&\!\! -\frac{\beta_2}{6}
\frac{d\pi_{\mu\nu}}{d\tau} + u_{\langle \mu; \nu \rangle}
\nonumber \\
&-& \frac{\beta^*_1}{6}(1-\kappa) \tilde{\phi} \pi_{\mu\nu}
+\alpha \Gamma \nabla_{\langle \mu} W_{\nu\rangle}
.\label{eq:GammaPhiPsi_1}
\end{eqnarray}
\end{subequations}
The arbitrary constant $\kappa \in (0,1)$ appearing in Eqs.
(\ref{eq:GammaPhiPsi_0}) and (\ref{eq:GammaPhiPsi_1}) is due to an
ambiguity originating from the last NexDC relation in
Eq.~(\ref{eq:pipi}) which allows us to write $\Pi^2=\kappa \Pi^2 +
(1-\kappa) \pi^{\mu\nu}\pi_{\mu\nu}/6$. To ensure that
$[su^{\mu}]_{;\mu}\ge 0$, we assume now (following the standard
2nd order theory \cite{MurongaPRC69}) a linear relationship
between the thermodynamic fluxes $\Pi, W^{\mu}, \pi^{\mu\nu}$ and
the corresponding thermodynamic forces, $\Upsilon, \Phi_{\mu},
\Psi_{\mu\nu}$,  with the usual transport coefficients $\zeta$,
$\lambda$, $\eta$, respectively. We have then
\begin{eqnarray}
  T [su^{\mu}]_{;\mu} = \frac{\Pi^2}{\zeta} -\frac{W^{\mu}W_{\mu}}{\lambda T}+\frac{\pi^{\mu\nu}\pi_{\mu\nu}}{2\eta}
\end{eqnarray}
and our $q$-hydrodynamics leads to the following constitutive
equations of the corresponding $d$-hydrodynamics:
\begin{subequations}
\label{eq:constitutive}
\begin{eqnarray}
\Pi \!\! &+& \!\! \tau_{\Pi}\frac{d\Pi}{d\tau} = -\zeta
u^{\mu}_{;\mu}
  -\kappa \beta^*_1 \zeta \tilde{\phi} \Pi + l_{\Pi W}\nabla_{\mu}W^{\mu}, \\
W_{\mu} \!\! &+& \!\! \tau_{W} \Delta_{\mu}^{\sigma} \frac{d
W_{\sigma}}{d\tau} =
 \lambda T \Big[ -(1+\alpha^*\Pi\Gamma) \frac{du_{\mu}}{d\tau} - \beta^*_1 \Pi\Gamma
 \phi_{\mu}  \nonumber \\
 &+& \frac{(\beta^*_0+\beta^*_1)}{6}
\frac{d\Gamma}{d\tau} W_{\mu} \Big]
 - l_{W\pi} \nabla_{\sigma}\pi^{\sigma}_{\mu}  -l_{W\Pi}\nabla_{\mu}\Pi ,   \\
\pi_{\mu\nu} \!\!&+&\!\! \tau_{\pi} \Delta_{\mu}^{\rho}
\Delta_{\nu}^{\sigma}\frac{d\pi_{\rho\sigma}}{d\tau}  = 2\eta  u_{\langle \mu;\nu\rangle}
\nonumber \\
 &-& \frac{\beta^*_1}{3} (1-\kappa) \eta  \tilde{\phi}\pi_{\mu\nu}
 +l_{\pi W}\nabla_{\langle \mu}W_{\nu\rangle},
\end{eqnarray}
\end{subequations}
with the corresponding relaxation times  $\tau_{\Pi} \equiv \zeta
(\beta^*_0 - \beta_2 )$, $\tau_{W} \equiv
\frac{1}{3}(\beta^*_1-\beta^*_0) \Gamma \lambda T$ and $\tau_{\pi}
\equiv\ \frac{1}{3}\beta_2 \eta$ and with the corresponding
heat-viscous coupling lengths $l_{\Pi W} =
\zeta(2\alpha-\alpha^*)\Gamma$, $l_{W\Pi}= 2\alpha \Gamma \lambda
T$, $l_{W\pi} =\alpha \Gamma \lambda T$ and $l_{\pi W}= 2\eta
\alpha \Gamma$. To ensure the positivity of relaxation times we
must have $\beta_0 > \beta_1+\beta_2$. Actually, one can also
express coefficients $\beta_{0,1,2}$ and $\alpha$  in terms of the
corresponding relaxation times and transport coefficients:
\begin{eqnarray}
&& \beta_0 = \frac{\tau_{\Pi}}{\zeta} +\frac{3\tau_{\pi}}{\eta}
 + \frac{3}{2}\frac{\tau_W}{\Gamma\lambda T},\quad
 \beta_1 = \frac{3}{2}\frac{\tau_W}{\Gamma\lambda T},
 \nonumber\\
 &&
\beta_2 = \frac{3\tau_{\pi}}{\eta},\quad
\alpha=\frac{l_{W\pi}}{\Gamma \lambda T}= \frac{l_{\pi W}}{2\Gamma\eta}.
\end{eqnarray}
To summarize this part, under the conditions mentioned above, and
with the help of the dissipative tensor relations,
Eq.~(\ref{eq:pipi}) and $q$-entropy conservation, one can obtain
the corresponding constitutive equations of $d$-hydrodynamics,
Eq.~(\ref{eq:constitutive}). They include the relaxation times and
heat-viscous coupling lengths in a quite natural way. Since the
original perfect $q$-hydrodynamics
Eq.~(\ref{eq:q-equation_of_motion}) does not contain any
space-time scale, it is than natural that the NexDC conjecture
does not introduce any definite relaxation time or viscous-heat
coupling length scale. However, as seen in
Eq.~(\ref{eq:constitutive}), the $d$-hydrodynamics obtained from
the $q$-hydrodynamics by using the NexDC relations, possesses  the
causal property of the corresponding $d$-hydrodynamics. Let us
close noticing that this $d$-hydrodynamics predicts that
\begin{eqnarray}
&&2l_{W\pi} = l_{W\Pi},\quad \frac{l_{W\pi}}{\lambda T}=
\frac{l_{\pi W}}{2\eta};\label{eq:testable1}\\
&&\frac{1}{\Pi/s} + \frac{ (l_{\Pi W} +\Gamma\tau_{\Pi}  )
}{\zeta/s} = \frac{( l_{\pi W}-3\Gamma\tau_{\pi})}{\eta/s} -
\frac{\tau_W}{\lambda T/s}.\qquad  \label{eq:testable2}.
\end{eqnarray}

\begin{figure}[h]
  \begin{center}
   \includegraphics[width=8.0cm]{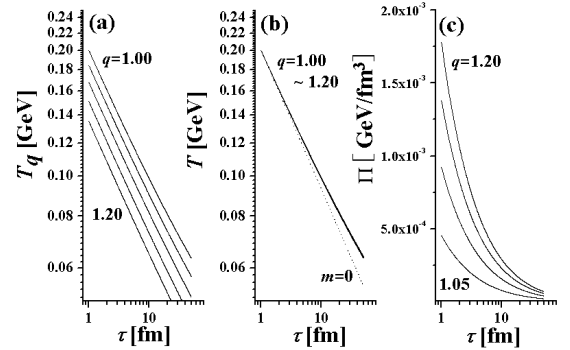}
   \caption{Proper time evolution of $T_q$, $T$ and $\Pi$
   for $q$= 1.00-1.20. The $q$-hydrodynamical
   evolution starts at $\tau_0 = 1.0$ fm. The initial
   temperatures, $T_q(\tau_0)$ are chosen in such way as to
   have $T_0 = 0.2$ GeV for all values of $q$.
   The doted line in (b) is obtained for a fluid
   composed of massless particles (in this case there is no $q$-dependence).
   }
   \label{Fig:1}
  \end{center}
\end{figure}

As an example, let us now apply the NexDC to a $q$-hydrodynamics
with the Bjorken type scaling initial conditions
\cite{BjorkenPRD27}. In this case the equation of motion
Eq.~(\ref{eq:q-equation_of_motion}) for an $q$-ideal fluid  has a
very simple form:
\begin{eqnarray}
 \frac{d
 \varepsilon_q(\tau)}{d\tau}=-\frac{\varepsilon_q(\tau)+P_q(\tau)}{\tau},
 \label{eq:q-hydro1+1}
\end{eqnarray}
with four velocity field $u_q^{\mu}=(1,0,0,0)$ (we use metric
$g_{\mu\nu}=(g_{\tau\tau},g_{xx},g_{yy},g_{\eta^*\eta^*})=(1,0,0,-\tau^2)$,
where $\tau=(t^2-z^2)^{1/2}$ and $\eta^*=\frac{1}{2} \ln
(t+z)/(t-z)$). The $q$-energy density and $q$-pressure are
defined, respectively, as $\varepsilon_q= u_{q\mu} {\cal
T}_q^{\mu\nu} u_{q\nu}$ and $P_q= -\frac{1}{3}{\cal T}^{\mu\nu}
\Delta_{q\mu\nu} $, by using the $q$-energy momentum tensor
\cite{OsadaPRC77}, ${\cal T}_q^{\mu\nu} \!\!= \!\!
\frac{1}{(2\pi\hbar)^3} \int \frac{d^3p}{p^0} [f_q(x,p)]^q$, with
the $f_q(x,p)$ given by Tsallis distribution,
\begin{eqnarray}
f_q(x,p)=\Big[ 1-(1-q)\frac{p_{\mu}u_q^{\mu}}{k_{\rm B}T_q(x)}\Big]^{1/(1-q)}.
\label{eq:q-distribution}
\end{eqnarray}
The corresponding EoS, $P_q=P_q(\varepsilon_q)$, is obtained by
eliminating the common parameter $T_q$ from both the $P_q(T_q)$
and $\varepsilon_q(T_q)$. Fig. \ref{Fig:1} shows the proper time
evolution of $T_q$, $T$ (notice that, because of Eq.
(\ref{eq:Nex/diss}), they are not independent) and the bulk
pressure $\Pi$ for the relativistic $\pi$ gas (with mass $m= $0.14
GeV) distributed according to Eq.~(\ref{eq:q-distribution}).
The $q$-hydrodynamical evolution starts at proper time $\tau_0 =
1.0$ fm with the $q$-temperature $T_q(\tau_0)$ chosen such that
$T_0  \equiv T(\tau_0)  = 0.2$ GeV. As seen in Fig.~\ref{Fig:1},
although $T_q$ depends on $q$, the  $T(\tau)$ does not (it scales
with $q$). This is interesting because in the usual
$d$-hydrodynamics cooling behavior (i.e., $T$) is affected by the
viscous effects. However, in our case the $q$-dependence of
Eq.~(\ref{eq:q-hydro1+1}) is only implicit and the corresponding
EoS depends only very weakly on $q$ in the high temperature region
(cf., Fig.~1 in \cite{OsadaPRC77}). On the other hand, it depends
on the mass of particles which composed our fluid. The bulk
pressure $\Pi$ decreases monotonously with $\tau$ and increases
with the non-extensive parameter $q$.

\begin{figure}[h]
  \begin{center}
   \includegraphics[width=8.0cm]{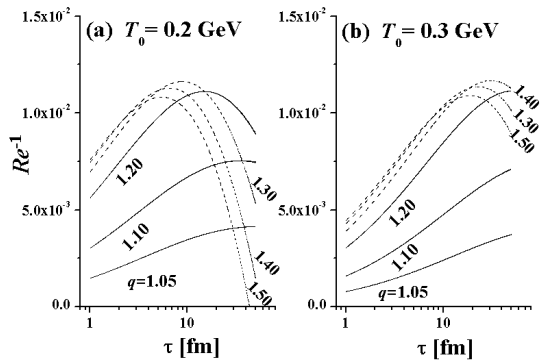}
   \caption{Proper time evolution of $R_e^{-1}$ obtained
   from $q$-hydrodynamics for two different initial temperatures (panels (a) and (b))
   and for different values of the parameter $q$. Solid lines are for, respectively,
   $q = 1.05$, $1.10$ and $1.20$, dashed lines are for $q = 1.30 - 1.50$. The same EoS
   has been used as in Fig.~\ref{Fig:1}.}
   \label{Fig:2}
  \end{center}
\end{figure}

We close this example by discussing the corresponding Reynolds
number \cite{BaymNPA418,KounoPRD41}. Separating in
Eq.~(\ref{eq:q-hydro1+1}) dissipative and non-dissipative terms by
using the Reynolds number $R_e$ \cite{MurongaPRC69} we get
\begin{eqnarray}
\frac{d\varepsilon}{d\tau} +\frac{\varepsilon+P}{\tau}
=\frac{(\varepsilon+P)}{\tau R_e}, \label{eq:1+1d-hydro}
\end{eqnarray}
In $q$-hydrodynamics with the NexDC conjecture
\begin{eqnarray}
 {R_e}^{-1}=\frac{3\tau}{\varepsilon+P} \left( \frac{\Pi}{\tau} + \frac{d\Pi}{d\tau} \right).
\label{eq:Reylords}
\end{eqnarray}
Fig.~\ref{Fig:2} shows examples of proper time evolution of the
inverse of Reynolds number obtained in Eq.~(\ref{eq:Reylords}).
For comparison, the maximal values of  $R_e^{-1}$ quoted in
\cite{BaierPRC73} for $\tau_0 = 1$ fm and $T_0 = 0.3$ GeV are
$R_e^{-1}\approx 0.065$ (for $\eta/s = 0.08$ and at $\tau \approx
2$ fm) and $R_e^{-1}\approx 0.21$ (for $\eta/s = 0.30$ and
$\tau\approx 4$ fm). Our $R_e^{-1}$ remains smaller than these
values, nevertheless we observe a similar maximum of $R_e^{-1}$ at
finite proper time, $\tau\sim 10$ fm. Notice that one can rewrite
Eq.~(\ref{eq:1+1d-hydro}) as $d \varepsilon/d\tau=(R^{-1}-1)
w/\tau$, which, in the case of $R^{-1}>1$, may indicate the
instability of our solution for large $\tau$ \cite{MurongaPRC69}.
However, as one can see in Fig.~\ref{Fig:2}, our
$q$-hydrodynamical model with $q$-scaling solution gives very
small values of $R_e^{-1}$,  $R_e^{-1}\ll 1$. This, according to
\cite{KounoPRD41,DenicolJPhysG35}, may guarantee the stability of
the corresponding $d$-hydrodynamics obtained.

To summarize: we have demonstrated that $d$-hydrodynamics obtained
from the perfect $q$-hydrodynamics (by means of the NexDC
conjecture) preserves its original causality. We believe then that
$q$-hydrodynamics can serve as a phenomenological model of
$d$-hydrodynamic with parameter $q$ describing summarily non
ideality of the hadronic fluid. We have also shown that the
corresponding inverse of the Reynolds number, $1/R_e$, may be
small enough to ensure a stable hydrodynamical evolution of the
corresponding strongly interacting quark/hadronic matter.

\begin{acknowledgments}
Partial support (GW) from the Ministry of Science and Higher
Education under contract 1P03B02230 is acknowledged.
\end{acknowledgments}

\end{document}